\documentclass{article}
\usepackage{theorem}
\usepackage{amssymb}
 \DeclareSymbolFont{lasy}{U}{lasy}{m}{n}
 \SetSymbolFont{lasy}{bold}{U}{lasy}{b}{n}
 \let\Box\undefined
 \DeclareMathSymbol\Box     {\mathord}{lasy}{"32}
\def\@abssec#1{\vspace{.05in}\footnotesize \parindent .2in
{\bf #1. }\ignorespaces}

\newtheorem{theorem}{Theorem}[section]
\newtheorem{lemma}[theorem]{Lemma}

\newcommand{\eps}{\varepsilon}

\newcommand{\dfrac}{\displaystyle\frac}

\newcommand{\pdr}[2]{\dfrac{\partial{#1}}{\partial{#2}}}

\newcommand{\commentout}[1]{}

\begin{document}

\title {Radiative transport limit for the random Schr\"odinger equation}
\author{Guillaume Bal \thanks{Department of Applied Physics and Applied Mathematics, Columbia University, New York NY, 10027; gbal@appmath.columbia.edu}
\and George
Papanicolaou \thanks{Department of Mathematics, Stanford CA, 94305;
papanico@math.stanford.edu}
\and Leonid Ryzhik \thanks{Department of
Mathematics, University of Chicago, Chicago IL, 60637;
ryzhik@math.uchicago.edu}
}

\maketitle
\begin{abstract} We give a detailed
mathematical analysis of the radiative transport limit
for the average phase space density of solutions of the Schr\"odinger
equation with time dependent random potential.
Our derivation is based on the construction of an approximate martingale
for the random Wigner distribution.
\end{abstract}

\section{Introduction}
\label{sec:intro}
The Schr\"odinger equation with random potential arises in many applications,
especially in wave propagation in random media, in the paraxial or parabolic
approximation. In this case the time harmonic wave field has the form
$u=e^{i\kappa z-\omega t}\phi(z,x)$ where $\kappa =\omega/c$ 
is the free space wave number, $z$ is the coordinate in the
direction of propagation, $x$ are the coordinates in the transverse directions
and $\phi$ satisfies the Schr\"odinger equation
\begin{equation}\label{eq:0}
2i\kappa \pdr{\phi}{z} + \Delta_x \phi + \kappa^2 \mu(z,x)\phi =0.
\end{equation}
Here $\mu(z,x)=n^2(z,x) -1$ denotes the fluctuations of the index of
refraction. The original scattering problem for
$(\Delta+\kappa^2n^2)u=0$ becomes an initial value problem for $\psi$
in $z$, in this approximation, so $\phi$ must be given at $z=0$.  The
validity of the parabolic approximation in random media under
different scaling limits is considered in \cite{jpf,PRS} and in
general in
\cite{parabolic}.

The purpose of this paper is to prove a theorem that establishes the
validity of the transport approximation for the average Wigner distribution
of $\phi$, in a suitable scaling limit and for a class of stochastic models
for the index of refraction fluctuations $\mu$ that are Markovian in $z$.
For this class of stochastic models it is possible to use martingale methods
to simplify the analysis.

Since the coordinate $z$ in the direction of propagation plays the role
of time in the parabolic approximation we will denote it by $t$ in the
rest of the paper. The problem then is to analyze the Schr\"odinger
equation with time dependent potential and to show that the associated
average Wigner distribution converges to the solution of a radiative
transport equation.

The propagation of wave energy in a scattering medium
is described phenomenologically by
radiative transport theory \cite{chandra} as follows.
Multiple scattering creates waves with all wave vectors
$k\in {\mathbb R}^d$ at every position $x\in {\mathbb R}^d$. Let us
denote by $W(t,x,k)$ the energy density of a wave having wave vector
$k$ at position $x$ at time $t$. The energy balance equation
has the form
\begin{equation}\label{rte}
\pdr {W(t,x,k)}t+k\cdot\nabla_xW(t,x,k)=
\int_{{\mathbb R}^d}dp~\sigma(x,k,p)W(t,x,p)-
\Sigma(x,k)W(t,x,k).
\end{equation}
Here $\sigma(x,k,p)$ is the probability to scatter from direction $p$
into direction $k$ at position $x$, and $\Sigma(x,k)$ is
the total probability to scatter from direction $k$ into some other
direction. When energy is conserved and scattering is symmetric,
\[
\Sigma(x,k)=\int_{{\mathbb R}^d}dp~\sigma(x,k,p),~~\sigma(x,k,p)=\sigma(x,p,k)
\]
equation (\ref{rte}) may be rewritten as
\begin{equation}\label{rte2}
\pdr {W(t,x,k)}t+k\cdot\nabla_xW(t,x,k)=
\int_{{\mathbb R}^d}dp~\sigma(x,k,p)[W(t,x,p)-W(t,x,k)].
\end{equation}

While various formal derivations of the radiative transport equation
(\ref{rte}), starting from the wave equation in a random medium, have
been known since the mid-1960's (see \cite{RPK-WM} for an extensive
bibliography) the mathematical methodology for doing this is not very
well developed.  A rigorous derivation of a spatially homogeneous
transport equation starting from the Schr\"odinger equation is given
by H. Spohn \cite{Spohn} who derived (\ref{rte}) for sufficiently
short times $t\in[0,t_0]$ and with a time independent Gaussian
potential. This result was extended to higher-order correlation
functions by T. Ho, L. Landau and A. Wilkins \cite{HLW} with the same
restrictions. Recently L. Erd\"os and H-T. Yau \cite{Erdos-Yau}
removed the small time restriction and considered more general initial
data. The idea in these proofs is to consider the Neumann series
expansion for the solution of the Schr\"odinger equation and to infer
appropriate estimates from it that allow passage to the limit.

In this paper we deal with time dependent random potentials for which
it is possible to analyze the transport approximations in a relatively
simple manner, without infinite Neumann expansions. We model the
random potential by a Markov process in time so that we can use
martingale methods and suitable test function expansions. A limit
theorem for one dimensional waves where such methods are used is given
in \cite{PW} and more general ones in \cite{BP}. An analysis of the
random Schr\"odinger equation with rapidly decorrelating in time
potential is given in \cite{PV}. Limit theorems for linear random
operator equations that decorrelate rapidly in time are given in \cite
{PV73}.  The random Schr\"odinger equation with delta function
potential is analyzed in \cite{DP} using martingale methods and its
equilibrium solutions are constructed in \cite{FPS}.

Formal derivations of radiative transport equations for various types
of waves in random media are given in \cite{BFPR,GW,RPK-WM}. Appendix
\ref{sec:append} contains such a derivation for the time-dependent
case. The results presented here extend to linear hyperbolic systems
with random coefficients such as those considered in \cite{RPK-WM}.
The transport equation for the Schr\"odinger equation with a
time-dependent potential was used recently in \cite{BPZ} to explain
pulse stabilization in time-reversal in a paraxial approximation to
the wave equation. The same phenomenon for general wave equations in
time-independent media was related to transport theory in
\cite{BR}.  Full justification of the results in \cite{BPZ} requires
analysis of the higher-order correlation functions and specific
scalings, as discussed in
\cite{PRS}.

{\bf Acknowledgment.} We thank K. Solna for numerous
discussions. G. Bal was supported by NSF grant DMS-0072008, L. Ryzhik
by NSF grant DMS-9971742 and G. Papanicolaou by  grants AFOSR F49620-01-1-0465 and
NSF-DMS-9971972.

\section{The main result}

\subsection{The Wigner distribution and the main result}

We consider the initial value problem  for the Schr\"odinger equation (\ref{eq:0})
in dimensionless form
\begin{equation}\label{eq:schroed}
i\eps\pdr{\phi_\eps}{t}+\frac{\eps^2}{2}\Delta\phi_\eps-
\sqrt{\eps}V\left(\frac{t}{\eps},\frac{x}{\eps}\right)\phi_\eps=0.
\end{equation}
The initial data $\phi_\eps^0(x)=\phi_\eps(0,x)$ are assumed to be
 uniformly bounded in $L^2({\mathbb R}^d)$:
\begin{equation}
  \label{eq:l2bds-init}
\|\phi_\eps^0\|_{L^2}\le C.
\end{equation}
This implies that
\begin{equation}
  \label{eq:l2bds}
\|\phi_\eps(t)\|_{L^2}=\|\phi_\eps^0\|_{L^2}\le C,~~ t\ge 0
\end{equation}
since the $L^2$-norm of the solution is preserved by the Schr\"odinger
evolution (\ref{eq:schroed}). Conservation of the $L^2$-norm implies
that for every realization of the random potential
$V_\eps(t,x)=V(t/\eps,x/\eps)$ there is a sequence $\eps_k\to 0$ so
that the energy density
\[
E_{\eps_k}(t,x)=|\phi_{\eps_k}(t,x)|^2
\]
has a weak limit $E(t,x)$ in $L^2([0,T]\times{\mathbb R}^d)$ as
$k\to\infty$.

It is well known that the limit $E(t,x)$ does not satisfy a closed
equation. A convenient way to study this limit is to consider energy
propagation in phase space that includes all positions $x$ and wave
vectors $k$ using the Wigner distribution $W_\eps(t,x,k)$, defined by
\begin{equation}\label{eq:wigdef}
W_\eps(t,x,k)=\int_{{\mathbb R}^d}\frac{dy}{(2\pi)^d}e^{ik\cdot
y}\phi_\eps(t,x-\frac{\eps y}{2}){\phi_\eps^*}(t,x+\frac{\eps
y}{2}).
\end{equation}
Here $*$ denotes complex conjugation. The basic properties of the
Wigner distribution can be found in
\cite{GMMP,LP}. In particular,
\[
\int dk~W_\eps(t,x,k)=|\phi_\eps(t,x)|^2=E_\eps(t,x)
\]
but $W_\eps$ may not be interpreted as energy density in phase space
since it is not necessarily positive. However, the limit of $W_\eps$
along a sub-sequence $\eps_k\to 0$ exists in ${\cal S}'({\mathbb
R}^d\times{\mathbb R}^d)$, the space of Schwartz distributions, and is
positive \cite{GMMP,LP}. We recall the proof of convergence in Section
\ref{sec:wigner}.  We say that the family $\phi_\eps^0(x)$ is pure if
the family of its Wigner distributions $W_\eps(x,k)$ converges weakly
as $\eps\to 0$ to a distribution $W_0(x,k)\in{\cal S}'$ without
restriction to a subsequence. We will assume that the initial data
form a pure family throughout the paper.

Our main result concerns the convergence of the expectation of the
Wigner distribution $W_\eps(t,x,k)$  defined by (\ref{eq:wigdef}) to
the solution of the radiative transport equation
(\ref{eq:treqformal}).
\begin{theorem}\label{thm1}
Let the random potential $V(t,x)$ satisfy the assumptions in Section
\ref{sec:random-potential} below. Let $W_0(x,k)$ be the limit Wigner measure
of the family $\phi_\eps^0(x)$, and let $\overline W(t,x,k)$ be the
weak probabilistic solution of the transport equation
\begin{equation}\label{eq:treqformal}
\pdr{\overline W}{t}+ k\cdot\nabla_x \overline W={\cal L}\overline W
\end{equation}
with initial data $W_0(x,k)$ and where 
the operator $\cal L$ is defined by
\begin{equation}\label{eq:cal-L}
{{\cal L}}\lambda=\int \frac{dp}{(2\pi)^d}
\hat R(\frac{p^2-k^2}{2},p-k)(\lambda(p)-\lambda(k)).
\end{equation}
Here
$\hat R(\omega,p)$ is the
Fourier transform of the correlation function of $V$, defined by
(\ref{eq:tildeR}). Then the expectation
$E\left\{W_\eps(t,x,k)\right\}$ of the Wigner distribution
$W_\eps(t,x,k)$ of the family $\phi_\eps(t,x)$ of solutions of
(\ref{eq:schroed}) converges weak-$*$ in $L^\infty([0,T];{\cal
S}'({\mathbb R}^d\times{\mathbb R}^d))$ to $\overline W(t,x,k)$.
\end{theorem}
By weak probabilistic solution we mean
that $\overline W(t,x,k)$ satisfies
\begin{equation}\label{eq:weak}
\langle \overline W,\lambda\rangle (t)-\langle W_0,\lambda|_{t=0}\rangle
=\int_0^t\Big\langle \overline W,\pdr{\lambda}{t}+k\cdot\nabla_x\lambda+{\cal
L}\lambda\Big\rangle (s)ds
\end{equation}
for all test functions $\lambda\in C^1([0,T];{\cal S})$. We actually
show that $E\left\{W_\eps(t,x,k)\right\}$ converges to $\overline W$
in a smaller space $L^\infty([0,T];{\cal A}'({\mathbb
R}^d\times{\mathbb R}^d))$.  The space ${\cal A}'$, which is defined
below in Section \ref{sec:wigner}, is more convenient for the analysis
of the Wigner distribution than ${\cal S}'$.

\subsection{The random potential}\label{sec:random-potential}

The random potential $V(t,x)$ is assumed to be stationary in space and
time and to have mean zero. It is constructed in  Fourier space as
follows. Let ${\cal V}$ be the set of measures of bounded
total variation with support inside a ball $B_L=\left\{|p|\le
L\right\}$
\begin{equation}\label{defcalv}
{\cal V}=\left\{\hat V:~\int_{{\mathbb R}^d}|d\hat V|\le C,
~\hbox{supp}~{\hat
V}\subset B_L,~ \hbox{$\hat V(p)=\hat V^*(-p)$}\right\}
\end{equation}
and let $\tilde V(t,p)$ be a mean-zero Markov process on $\cal V$ with
generator $Q$. The time-dependent random potential $V(t,x)$ is given by
\[
V(t,x)=\int \frac{d\tilde V(t,p)}{(2\pi)^d}e^{ip\cdot x}
\]
and is real and uniformly bounded:
\[
|V(t,x)|\le C.
\]
We assume that the process $V(t,x)$ is stationary in $t$ and $x$
with correlation function $R(t,x)$
\[
E\left\{V(t,x)V(t+s,x+z)\right\}=R(s,z)~~~\hbox{for all
$x,z\in{\mathbb R}^d$, and $t,s\in{\mathbb R}$.}
\]
In terms of the process $\tilde V(t,p)$ this means that given any two
bounded continuous functions $\hat\phi(p)$ and $\hat\psi(p)$ we have
\begin{equation}\label{powerspectrum}
E\left\{\langle \tilde V(s),\hat\phi\rangle\langle
\tilde V(t+s),\hat\psi\rangle\right\}=
(2\pi)^d\int dp \tilde R(t,p)\phi(p)\hat\psi(-p).
\end{equation}
Here $\langle\cdot,\cdot\rangle$ is the usual duality product on
${\mathbb R}^d\times{\mathbb R}^d$, and the power spectrum $\tilde R$
is the Fourier transform of $R(t,x)$ in $x$:
\[
\tilde R(t,p)=\int dx e^{-ip\cdot x}R(t,x).
\]
We assume that $\tilde R(t,p)\in {\cal S}({\mathbb
R}\times{\mathbb R}^d)$ for simplicity and define $\hat R(\omega,p)$
as
\begin{equation}\label{eq:tildeR}
\hat R(\omega,p)=\int dt e^{-i\omega t}\tilde R(t,p),
\end{equation}
which is the space-time Fourier transform of $R$.
 
We assume that the generator $Q$ is a bounded operator on
$L^\infty({\cal V})$ with a unique invariant measure $\pi(\hat V)$
\[
Q^*\pi=0.
\]
and that there exists $\alpha>0$ such that if $\langle g,\pi\rangle=0$
then
\begin{equation}\label{eq:expdecaynew}
\|e^{rQ}g\|_{L_{\cal V}^\infty}\le C\|g\|_{L_{\cal
V}^\infty}e^{-\alpha t}.
\end{equation}
The simplest example of a generator with gap in the spectrum and
invariant measure $\pi$ is a jump process on ${\cal V}$ where
\[
Qg(\hat V)=\int_{\cal V}g (\hat V_1)d\pi(\hat V_1)-g(\hat V),~~~
\int_{\cal V}d\pi(\hat V)=1.
\]
Given (\ref{eq:expdecaynew}), the Fredholm alternative holds for the
Poisson equation
\[
Qf=g,
\]
provided that $g$ satisfies $\langle\pi,g\rangle=0$.  It has a unique
solution $f$ with $\langle\pi,f\rangle=0$ and $\|f\|_{L^\infty_{V}}\le
C\|g\|_{L^\infty_{V}}$. The solution $f$ is given explicitly by
\[
f(\hat V)=-\int_0^\infty dr e^{rQ}g(\hat V),
\]
and the integral converges absolutely because of
(\ref{eq:expdecaynew}).

\subsection{General convergence of the Wigner distribution}\label{sec:wigner}

Existence of the limit of the Wigner family $W_\eps(t,x,k)$ defined by
(\ref{eq:wigdef}) is shown as follows.  We introduce the space ${\cal
A}$, as in \cite{LP}, of functions $\lambda(x,k)$ of $x$ and $k$ such
that
\[
\|\lambda\|_{{\cal A}}=\int_{{\mathbb R}^{2d}}dy\sup_{x}
|\tilde \lambda(x,y)|<\infty,
\]
where
\begin{equation}\label{ftk}
\tilde\lambda(x,y)=\int_{{\mathbb R}^{d}}
dk e^{-ik\cdot y}\lambda(x,k)
\end{equation}
is the Fourier transform of $\lambda$ in $k$. Convergence in the space
${\cal A}$ is easier to establish than in ${\cal S}$ because its
definition does not involve derivatives. Moreover, as the following lemma
shows, the distributions $W_\eps$ are uniformly bounded in ${\cal A}'$,
the dual space to ${\cal A}$. 
\begin{lemma}\label{lemma1}\cite{LP}
The family $W_\eps(t,x,k)$ is uniformly bounded in ${\cal A}'$, that
is, there exists a constant $C>0$ independent of $t$ so that:
\begin{equation}
\label{eq:boundWeps}
\|W_\eps(t)\|_{{\cal A}'}\le C
\end{equation}
for all $\eps>0$ and $t\ge 0$.
\end{lemma}
{\bf Proof.} Let $\lambda(x,k)\in{\cal A}$. Then, 
\begin{eqnarray*}
&&
\langle W_\eps(t),\lambda\rangle=\int dxdk W_\eps(t,x,k)\lambda(x,k)=
\int\frac{dxdkdy}{(2\pi)^d}e^{ik\cdot y}\phi_\eps(t,x-\frac{\eps y}{2})
\phi_\eps^*(t,x+\frac{\eps y}{2})\lambda(x,k)\\
&&~~~~~~~~~~~~~~~=\int\frac{dxdy}{(2\pi)^{d}}
\phi_\eps(t,x+\frac{\eps y}{2})
\phi_\eps^*(t,x-\frac{\eps y}{2})\tilde\lambda(x,y).
\end{eqnarray*}
Therefore, using the Cauchy-Schwartz inequality in $x$
\[
\left|\langle W_\eps(t),\lambda\rangle\right|\le
\int\frac{dy dx}{(2\pi)^{2d}}\left|\tilde\lambda(x,y)\right|
\left|\phi_\eps(t,x-\frac{\eps y}{2})\right|
\left|\phi_\eps(t,x+\frac{\eps y}{2})\right|
\le
\|\phi_\eps(t)\|_{L_2}^2\|\lambda\|_{{\cal A}}\le C\|\lambda\|_{{\cal
A}},
\]
where we use the conservation of the $L^2$-norm (\ref{eq:l2bds}) in the
last step. This gives (\ref{eq:boundWeps}).

Lemma \ref{lemma1} implies that at every time $t\ge 0$ we can choose a
sequence $\eps_j\to 0$ so that $W_{\eps_j}$ converges weakly in ${\cal
A}'\subset{\cal S}'$ to a limit distribution $W(t)$. One can show
\cite{GMMP} that the limit measure $W(t)$ is non-negative and may thus
be interpreted as the limit energy density in phase space. Moreover,
if there are no oscillations in the initial data on scales smaller
than $\eps$ then the limit captures correctly the behavior of the
energy $E_\eps(t,x)$. More precisely, if
\begin{equation}\label{eps-oscill}
\eps\|\nabla\phi_\eps^0\|_{L_x^2}\le C
\end{equation}
then for any test function $\theta(x)\in {\cal S}({\mathbb R}^d)$ we have
\[
\int dxdk\theta(x)W(t,x,k)=\lim_{\eps\to 0}\int
dx\theta(x)|\phi_\eps(t,x)|^2.
\]
Condition (\ref{eps-oscill}) is sufficient but not necessary for this.

\section{Convergence of the expectation}\label{sec:expect}

We prove Theorem \ref{thm1} in this section. 
The proof
is based on the method of \cite{PW} and proceeds as
follows. The distribution $W_\eps$ defined by (\ref{eq:wigdef}) satisfies 
\begin{eqnarray}\label{eq:wigeqnew}
&&\pdr{W_\eps}{t}+k\cdot\nabla_xW_\eps=\frac 1{i\sqrt{\eps}}
\int \frac{d\tilde V(t/\eps,p)}{(2\pi)^d} e^{ip\cdot x/\eps}
\left[W_\eps(t,x,k-\frac{p}{2})-W_\eps(t,x,k+\frac{p}{2})\right]\\
&&W_\eps(0,x,k)=W_\eps^0(x,k).\nonumber
\end{eqnarray}
Here $W_\eps^0$ is the Wigner distribution of the family
$\phi_\eps^0(x)$, the initial data for (\ref{eq:schroed}). 
The Cauchy
problem (\ref{eq:wigeqnew}) generates a measure $P_\eps$ on the space
$C([0,T];{\cal A}')$ of continuous functions in time with values in
${\cal A}'$. It is supported on paths inside a ball
$X=\left\{W\in{\cal A}':~\|W\|_{{\cal A}'}\le C\right\}$ with the
constant $C$ as in (\ref{eq:boundWeps}). The set $X$ is the state
space for the random process $W_\eps(t)$. The joint process $(\tilde
V(t/\eps),W_\eps(t))$ takes values in the space ${\cal V}\times X$.  We
will denote by $\tilde P_\eps$ the corresponding measure on the space
${\cal V}\times X$ generated by (\ref{eq:wigeqnew}) and the process
$\tilde V(t/\eps)$.  
Let us fix a deterministic function
$\lambda\in C^1([0,T];{\cal S})$.  We will show that the functional
$G_\lambda:~C([0,T];X)\to C[0,T]$ defined by
\[
G_\lambda[W](t)=\langle W,\lambda\rangle(t)-\int_0^t\langle
W,\pdr{\lambda}{t}+k\cdot\nabla_x\lambda+{\cal L}\lambda\rangle(s)ds
\]
is an approximate $P_\eps$-martingale. More precisely, we will show that
\begin{equation}\label{eq:approxmart}
\left|E^{P_\eps}\left\{G_\lambda[W](t)|{\cal F}_s\right\}
-G_\lambda[W](s)\right|\le
C_{\lambda,T}\sqrt{\eps}
\end{equation}
uniformly for all $W\in C([0,T];X)$ and $0\le s<t\le T$. 

The next step
is to show that the measures $P_\eps$ form a tight family, so that
there exists a subsequence $\eps_j\to 0$ so that $P_\eps$ converges
weakly to a measure $P$ supported on $C([0,T];X)$. Weak convergence of
$P_\eps$ and the strong convergence (\ref{eq:approxmart}) together imply
that $G_\lambda[W](t)$ is a $P$-martingale so that
\begin{equation}\label{eq:Pmart}
E^{P}\left\{G_\lambda[W](t)|{\cal F}_s\right\}-G_\lambda[W](s)=0.
\end{equation}
Taking $s=0$ in this we obtain the transport equation
(\ref{eq:treqformal}) for $\overline W=E^P\left\{W(t)\right\}$, in its
weak formulation (\ref{eq:weak}). 

The limit measure $P$ may depend on
the choice of the subsequence $\eps_j\to 0$ but the expectation
$\overline W$ being the unique solution of (\ref{eq:treqformal}) does
not depend on it. Therefore the whole family $E\left\{W_\eps\right\}$
converges to $\overline W$ as $\eps\to 0$ in $L^\infty([0,T];{\cal
S}')$.  Furthermore, the a priori bound (\ref{eq:boundWeps}) implies
that $W_\eps(t,x,k)$ converges weak-$*$ in $L^\infty([0,T];{\cal A}')$
for every realization of the random potential. Therefore the result
above implies that actually $E\left\{W_\eps\right\}$ converges to
$\overline W$ in $L^\infty([0,T];{\cal A}')$.

\commentout{
The
uniformity of convergence in ${\cal S}'({\mathbb R}^d\times{\mathbb
R}^d)$ is shown as follows. Let
$Z_\eps=E\left\{W_\eps\right\}-\overline W$, then we have from
(\ref{eq:approxmart})
\begin{equation}
\left|\langle Z_\eps,\lambda\rangle(t)-
\int_0^t \langle Z_\eps,k\cdot\nabla_x\lambda+{\cal L}\lambda\rangle(s)ds\right|\le C\sqrt{\eps}.
\end{equation}
The operator $k\cdot\nabla_x+{\cal L}$ is a bounded operator on ${\cal
S}({\mathbb R}^d\times{\mathbb R}^d)$, therefore we obtain
\[
\left|\langle Z_\eps,\lambda\rangle(t)\right|\le
C\int_0^t\|Z_\eps\|_{{\cal S}'}(s)ds\|\lambda\|_{\cal S}+C\sqrt{\eps}.
\]
Thus we have
\[
\|Z_\eps\|_{{\cal S}'}(t)\le C\int_0^t\|Z_\eps\|_{{\cal S}'}(s)ds +C_\lambda\sqrt{\eps}
\]
and hence $\|Z_\eps\|_{{\cal S}'}(t)\to 0$ uniformly on finite time intervals as follows
from the Gronwall's inequality.
}
The proof is organized as follows.
The approximate martingale property (\ref{eq:approxmart}) is proved in
Sections \ref{sec:constr} and \ref{sec:bounds}. The weak compactness
of the family $P_\eps$ is proved in Section \ref{sec:compact}.

\subsection{Construction of the test functions}\label{sec:constr}

In order to obtain the approximate martingale property
(\ref{eq:approxmart}) one has to consider conditional expectation of
functions $F(W,\hat V)$.  The only functions we will need to consider
are those of the form $F(W,\hat V)=\langle W,\lambda(\hat V)\rangle$
with $\lambda\in L^\infty({\cal V};C^1([0,T];{\cal S}({\mathbb
R}^d\times{\mathbb R}^d)))$. Given a function $F(W,\hat V)$ let us
define the conditional expectation
\[
E_{W,\hat V,t}^{\tilde P_\eps}\left\{F(W,\hat V)\right\}(\tau)=
E^{\tilde P_\eps}\left\{F(W(\tau),\tilde V(\tau))|~W(t)=W, \tilde V(t)=\hat V\right\},~~ \tau\ge t.
\]
The weak form of the infinitesimal generator of the Markov process
generated by $\tilde P_\eps$ is given by
\begin{equation}
\label{generator}
\left.\frac{d}{dh}E_{W,\hat V,t}^{\tilde P_\eps}\left\{\langle
W,\lambda(V)\rangle\right\}(t+h)\right|_{h=0}= \frac 1\eps \langle
W,Q\lambda\rangle+\left\langle
W,\left(\pdr{}{t}+k\cdot\nabla_x+\frac{1}{\sqrt{\eps}}{\cal K}[\hat V,\frac
x\eps]\right)\lambda\right\rangle
\end{equation}
and hence
\[
G_\lambda^\eps=\langle W,\lambda(V)\rangle(t)-\int_0^t\left\langle
W,\left(\frac 1\eps Q+ \pdr{}{t}+k\cdot\nabla_x+\frac{1}{\sqrt{\eps}}{\cal
K}[\hat V,\frac x\eps]\right)\lambda\right\rangle(s)ds
\]
is a $\tilde P_\eps$-martingale. 
The operator ${\cal K}$ is defined by
\begin{equation}
  \label{eq:Koper}
{\cal K}[\hat V,z]\psi(x,z,k,\hat V)=
\frac 1i\int \frac{d\hat V(p)}{(2\pi)^d}
e^{ip\cdot z}
\left[\psi(x,z,k-\frac p2)-\psi(x,z,k+\frac p2)\right].
\end{equation}
The generator (\ref{generator}) comes from equation (\ref{eq:wigeqnew}) 
written in the form
\begin{equation}
\label{eq:wigner2}
\partial_t W_\eps + k\cdot\nabla_x W_\eps=\frac{1}{\sqrt{\eps}}
{\cal K}[\tilde V(t/\eps),x/\eps] W_\eps.
\end{equation}

Given a test function $\lambda(t,x,k)\in C^1([0,T];{\cal S})$ we 
construct a function
\begin{equation}\label{eq:lambdaapprox}
\lambda_\eps(t,x,k,\hat V)=\lambda(t,x,k)+
\sqrt{\eps}\lambda_1^\eps(t,x,k,\hat V)+
\eps\lambda_2^\eps(t,x,k,\hat V)
\end{equation}
with $\lambda_{1,2}^\eps(t)$ bounded in $L^\infty({\cal V};{\cal
A}({\mathbb R}^d\times{\mathbb R}^d))$ uniformly in $t\in[0,T]$. It is
sufficient for us to prove the simpler bound for the correctors in
${\cal A}$ instead of ${\cal S}$ because of the a priori bound
(\ref{eq:boundWeps}) for $W_\eps$ in ${\cal A}'$.  The functions
$\lambda_{1,2}^\eps$ will be chosen so that
\[
\|G_{\lambda_\eps}^\eps(t)-G_\lambda(t)\|_{L^\infty({\cal V})}\le
C_\lambda\sqrt{\eps}
\]
for all $t\in[0,T]$.  The approximate martingale property
(\ref{eq:approxmart}) follows from this.  The approximate test
function $\lambda_\eps(t,x,k)$ in (\ref{eq:lambdaapprox}) is constructed in a
manner similar to the formal asymptotic expansion (\ref{eq:wexp})
considered in Appendix \ref{sec:append}. 

The functions
$\lambda_1^\eps$ and $\lambda_2^\eps$ are as follows.  Let
$\lambda_1(t,x,z,k,\hat V)$ be the mean-zero solution of the Poisson equation
\begin{equation}\label{lambda1eq}
k\cdot\nabla_z\lambda_1+Q\lambda_1=-{\cal K}\lambda.
\end{equation}
It is given explicitly by
\[
\lambda_1(t,x,z,k,\hat V)=\frac 1i\int_0^\infty dr e^{rQ}
\int \frac{d\hat V(p)}{(2\pi)^d}
e^{ir(k\cdot p)+i(z\cdot p)}
\left[\lambda(t,x,k-\frac p2)-\lambda(t,x,k+\frac p2)\right].
\]
Then we let
$\lambda_1^\eps(t,x,k,\hat V)=\lambda_1(t,x,x/\eps,k,\hat V)$. The second order
corrector is 
$\lambda_2^\eps(t,x,k,\hat V)=\lambda_2(t,x,x/\eps,k,\hat V)$ where
$\lambda_2(t,x,z,k,\hat V)$ is the mean-zero solution of
\begin{equation}\label{lambda2eq}
k\cdot\nabla_z\lambda_2+Q\lambda_2=
{ {\cal L}}\lambda-{\cal K}\lambda_1,
\end{equation}
which exists because $E\left\{{\cal K}\lambda_1\right\}={ {\cal L}}\lambda$,
and is given by
\begin{eqnarray*}
&&\lambda_2(t,x,z,k,\hat V)=-\int_0^\infty dr e^{rQ}
\left[{\cal L}\lambda(t,x,k)-[{\cal K}\lambda_1](t,x,z+rk,k,\hat V)\right].
\end{eqnarray*}
Using (\ref{lambda1eq}) and
(\ref{lambda2eq}) we have
\begin{eqnarray*} 
&&\mbox{}\!\!\!\!\!\!\!\!\!\!\!\!
\left.\frac{d}{dh}E_{W,\hat V,t}^{\tilde P_\eps}
\left\{\langle W,{\lambda_\eps}\rangle\right\}(t+h)\right|_{h=0}
=\left\langle W,\left(\pdr{}{t}+k\cdot\nabla_x+
\frac{1}{\sqrt{\eps}}{\cal K}[\hat V,\frac x\eps]+
\frac 1\eps Q\right)
\left(\lambda+\sqrt{\eps}\lambda_1^\eps+\eps\lambda_2^\eps\right)\right\rangle
\\
&&=\left\langle W, \left(\pdr{}{t}+k\cdot\nabla_x\right)
  \lambda+{\cal L}\lambda\right\rangle
+\left\langle W, 
   \left(\pdr{}{t}+k\cdot\nabla_x\right)\left(\sqrt{\eps}\lambda_1^\eps+
\eps\lambda_2^\eps\right)
+
\sqrt{\eps}{\cal K}[\hat V,\frac x\eps]\lambda_2^\eps\right\rangle\\
&&=\left\langle W,\left(\pdr{}{t}+k\cdot\nabla_x\right)\lambda+{\cal
L}\lambda\right\rangle+\sqrt{\eps}\langle W,\zeta_\eps^\lambda\rangle
\end{eqnarray*}
with
\[
\zeta_\eps^\lambda=\sqrt{\eps}
\left(\pdr{}{t}+k\cdot\nabla_x\right)\lambda_1^\eps+
\eps \left(\pdr{}{t}+k\cdot\nabla_x\right)\lambda_2^\eps+
\sqrt{\eps}{\cal K}[\hat V,\frac x\eps]\lambda_2^\eps.
\]
The terms $k\cdot\nabla_x\lambda_{1,2}^\eps$ above are understood as
differentiation with respect to the slow variable $x$ only, and not
with respect to $x/\eps$. It follows that $G_{\lambda_\eps}^\eps$
is given by
\[
G_{\lambda_\eps}^\eps(t)=\langle W(t),\lambda_\eps\rangle-
\int_0^t ds\left\langle W,\left(\pdr{}{t}+k\cdot\nabla_x+{\cal L}\right)
\lambda\right\rangle(s)-\sqrt{\eps}\int_0^t ds\langle W,\zeta_\eps^\lambda\rangle(s)
\]
and is a martingale with respect to the measure $\tilde P_\eps$
defined on $D([0,T];X\times{\cal V})$, the space of right-continuous
paths with left-side limits \cite{billingsley1}. The estimate
(\ref{eq:approxmart}) follows from the following two lemmas.
\begin{lemma}\label{lemma1new}
Let $\lambda\in C^1([0,T];{\cal S}({\mathbb R}^d\times{\mathbb
R}^d))$. Then there exists a constant $C_\lambda>0$ independent of time
$t\in[0,T]$ so that the
correctors $\lambda_1^\eps(t)$ and $\lambda_2^\eps(t)$ satisfy the uniform
bounds
\begin{equation}\label{eq:lambda1bdnew}
\|\lambda_1^\eps(t)\|_{{L^\infty({\cal V};{\cal A})}}+
\|\lambda_2^\eps(t)\|_{{L^\infty({\cal V};{\cal A})}}\le C_\lambda
\end{equation}
and
\begin{equation}\label{eq:lambda2bdnew}
\Big\|\pdr{\lambda_1^\eps(t)}{t}+
k\cdot\nabla_x\lambda_1^\eps(t)\Big\|_{{L^\infty({\cal V};{\cal A})}}+
\Big\|\pdr{\lambda_2^\eps(t)}{t}+
k\cdot\nabla_x\lambda_2^\eps(t)\Big\|_{{L^\infty({\cal V};{\cal
A})}}\le C_\lambda.
\end{equation}
\end{lemma}
\begin{lemma}\label{lemma3}
There exists a constant $C_\lambda$ such that
\[
\|{\cal K}[\hat V,x/\eps]\|_{{\cal A}\to{\cal A}} \le C
\]
for any $\hat V\in{\cal V}$ and all $\eps\in(0,1]$.
\end{lemma}

Indeed, (\ref{eq:lambda1bdnew}) implies that $\left|\langle W,\lambda\rangle
- \langle W,\lambda_\eps\rangle\right|\le C\sqrt{\eps}$ for
all $W\in X$ and $V\in{\cal V}$, while (\ref{eq:lambda2bdnew}) and
Lemma \ref{lemma3} imply that for all $t\in[0,T]$
\begin{equation}\label{zetabd}
\|\zeta_\eps^\lambda(t)\|_{\cal A}\le C
\end{equation}
for all $V\in{\cal V}$ so that (\ref{eq:approxmart}) follows.

{\bf Proof of Lemma \ref{lemma3}.}  Let $g_\eps(x,k)={\cal
K}[V,x/\eps]\eta(x,k)$ with $\eta(x,k)\in{\cal A}$. Then
\begin{eqnarray*}
\tilde g_\eps(x,y)=\int \frac{d\hat V(p)}{(2\pi)^d} e^{ip\cdot x/\eps}
\tilde\eta(x,y)\left[e^{-ip\cdot y/2}-e^{ip\cdot y/2}\right]
\end{eqnarray*}
and thus
\[
|\tilde g_\eps(x,y)|\le C|\tilde\eta(x,y)|
\]
and the conclusion of Lemma \ref{lemma3} follows.

\subsection{Bounds on the correctors}\label{sec:bounds}

We now prove Lemma \ref{lemma1new}. We will omit the time dependence
of the test function $\lambda$ to simplify the
notation.

{\bf Proof of Lemma \ref{lemma1new}.} We first prove
(\ref{eq:lambda1bdnew}). The Fourier transform of
$\lambda_1^\eps$ in $k$ is given by
\begin{equation}\label{tildelambda1}
\tilde\lambda_1^\eps(x,y,\hat V)=
\frac 1i\int_0^\infty dr e^{rQ}\int \frac{d\hat V(p)}{(2\pi)^d}
\tilde\lambda(x,y-rp)
e^{ix/\eps\cdot p} \left[e^{-ip\cdot (y-rp)/2}-
e^{ip\cdot (y-rp)/2}\right].
\end{equation}
Therefore using (\ref{eq:expdecaynew}) we obtain
\begin{equation}\label{linftybdtilde1}
\|\tilde\lambda_1^\eps(x,y,\hat V)\|_{L_{x,y}^\infty}\le
\frac{C}{\alpha}\|\tilde\lambda\|_{L_{x,y}^\infty}
\end{equation}
uniformly for all $\hat V\in{\cal V}$. It is therefore sufficient to consider
$|y|>2$.  Let  $S(y)=(|y|-1)/4L$ with $L$ as in the definition
(\ref{defcalv}) of the set ${\cal V}$. We write (\ref{tildelambda1}) as
\begin{eqnarray*}
\tilde\lambda_1^\eps(x,y,\hat V)=J_{r<S(y)}+J_{r>S(y)}
\end{eqnarray*}
with
\[
J_{r<S(y)}=\frac 1i\int_0^{S(y)}dr e^{rQ}\int\frac{d\hat V(p)}{(2\pi)^d}
\tilde\lambda(x,y-rp)
e^{ix/\eps\cdot p} \left[e^{-ip\cdot (y-rp)/2}-
e^{ip\cdot (y-rp)/2}\right]
\]
and
\[
J_{r>S(y)}=\frac 1i\int_{S(y)}^{\infty}
dr e^{rQ}\int\frac{d\hat V(p)}{(2\pi)^d}
\tilde\lambda(x,y-rp)
e^{ix/\eps\cdot p} \left[e^{-ip\cdot (y-rp)/2}-
e^{ip\cdot (y-rp)/2}\right].
\]
We estimate each of these two terms separately.

To bound $J_{r<S(y)}^1$ we note that since $\lambda$ is of the
Schwartz class we have for $|p|\le L$ and $r<S(y)$
\begin{eqnarray*}
&&\left|\tilde\lambda(x,y-rp)\right|\le
\sup_{|z-y|\le rL}\left|\tilde\lambda(x,z)\right|\le
\sup_{|z|\ge{|y|/2}}\left|\tilde\lambda(x,z)\right|\le
\frac{C_\lambda}{|y|^{5d}}.
\end{eqnarray*}
Then we obtain
\begin{eqnarray}
&&\int_{|y|\ge 2}dy
\sup_{x,V}\left|J_{r<S(y)}(x,y,\hat V)\right|\le
C_\lambda\int_{|y|\ge 2} \frac{dy}{|y|^{5d}}\int_0^{S(y)}dr
e^{-\alpha r}\le C_{\lambda,\alpha}.
\label{phiest}
\end{eqnarray}
Next we note that
\begin{equation}\label{IrRy}
\int_{|y|\ge 2} dy
\sup_{x,\hat V}|J_{r>S(y)}(x,y,\hat V)|\le\|\tilde\lambda\|_{L_{x,y}^\infty}
\int_{|y|\ge 2} dy\int_{S(y)}^\infty dre^{-\alpha r}
\le C_\alpha\|\tilde\lambda\|_{L_{x,y}^\infty}.
\end{equation}
Therefore (\ref{linftybdtilde1}), (\ref{phiest})
and (\ref{IrRy}) imply that
\begin{equation}\label{lambda1tildebd}
\|\lambda_1^\eps\|_{\cal A}\le C_{\lambda,\alpha}
\end{equation}
for all $\hat V\in{\cal V}$. 

We show next that $\lambda_2^\eps$ is uniformly bounded. This is done in
several steps that we formulate as separate lemmas. Define
\[
I(x,y)=\int_0^\infty dr e^{-\alpha r}\sup_{\hat V}
\int \frac{|d\hat V(p)|}{(2\pi)^d}
\int_0^\infty ds e^{-\alpha s}\sup_{\hat V_1}
\int \frac{|d\hat V_1(q)|}{(2\pi)^d}
|\tilde\lambda(x,y-rp-(r+s)q)|.
\]
\begin{lemma}\label{sublemma1}
We have the estimate
\begin{eqnarray}\label{eq:sublemma1}
|\tilde\lambda_2^\eps(x,y,\hat V)|\le C_\alpha\left[I(x,y)+
|\widetilde{{\cal L}\lambda}(x,y)|\right]
\end{eqnarray}
\end{lemma}
\begin{lemma}\label{sublemma2}
For the limit 
 operator ${\cal L}$ we have the bound
\begin{equation}\label{bdforlbar}
\| {\cal L}\lambda\|_{\cal A}\le \int d\xi dy|
\widehat{ {\cal L}\lambda}(\xi,y)|\le C\int d\xi dy|\hat \lambda(\xi,y)|.
\end{equation}
Here $\hat f$ denotes the Fourier transform of $f$ both in $x$ and $k$.
\end{lemma}
We split $I(x,y)$ as
\begin{equation}\label{eq:i=i<+i>}
I(x,y)=I_{r<S(y)}+I_{r>S(y)}=I_{r<S(y)}^{s<S(y)}+
I_{r<S(y)}^{s>S(y)}+
I_{r>S(y)}.
\end{equation}
\begin{lemma}\label{sublemma3}
We have the following bounds:
\begin{equation}
\int dy\sup_{x}\left[I_{r>S(y)}(x,y)+I_{r<S(y)}^{s>S(y)}\right]\le
C_\alpha\|\tilde\lambda\|_{L_{x,y}^\infty}
\end{equation}
and
\begin{equation}
\int dy\sup_{x}I_{r<S(y)}^{s<S(y)}\le C_\alpha\sup_{x\in{\mathbb R}^d,
|y|\ge 1}\left\{|y|^{5d}
|\tilde\lambda(x,y)|\right\}.
\end{equation}
\end{lemma}
Lemmas \ref{sublemma1}, \ref{sublemma2} and \ref{sublemma3} imply
clearly that $\|\lambda_\eps^2\|_{\cal A}\le C$ for all $\hat V\in{\cal V}$.
This finishes the proof of (\ref{eq:lambda1bdnew}). The proof of
(\ref{eq:lambda2bdnew}) is quite similar and is therefore omitted.

We now prove Lemmas \ref{sublemma1}-\ref{sublemma3} to conclude the
proof of Lemma \ref{lemma1new}.

{\bf Proof of Lemma \ref{sublemma1}.} The
Fourier transform of $\lambda_2^\eps$ in $k$ is given by
\begin{eqnarray*}
&&\tilde\lambda_2^\eps(x,y,\hat V)=-\int_0^\infty dr e^{rQ}
 \int dk e^{-ik\cdot y}
\left[{\cal L}\lambda(x,k)-
\frac 1i\int \frac{d\hat V(p)}{(2\pi)^d}e^{ip\cdot(x/\eps+rk)}\right.\\
&&\left.~~~~~~~~~~~~~~~
\times\left[\lambda_1(x,\frac x\eps+rk,k-\frac p2,\hat V)-
\lambda_1(x,\frac {x}{\eps}+rk,k+\frac p2,\hat V)\right]\right].
\end{eqnarray*}
The second term above may be written as
\begin{eqnarray*}
&&\frac 1i\int dk e^{-ik\cdot y}
\int\frac{d\hat V(p)}{(2\pi)^d}
e^{ip\cdot(x/\eps+rk)}
\left[\lambda_1(x,\frac x\eps+rk,k-\frac p2,\hat V)-
\lambda_1(x,\frac {x}{\eps}+rk,k+\frac p2,\hat V)\right]\\
&&=-\int dk e^{-ik\cdot y}\int d\hat V(p)e^{ip\cdot(x/\eps+rk)}
\int_0^\infty ds e^{sQ}\int \frac{d\hat V(q)}{(2\pi)^d}
e^{is(k-p/2)\cdot q+i(x/\eps+rk)\cdot q}\\
&&\times
\left[\lambda(k-\frac p2-\frac q2)-\lambda(k-\frac p2+\frac q2)\right]
\\
&&+
\int dk e^{-ik\cdot y}\int \frac{d\hat V(p)}{(2\pi)^d}
e^{ip\cdot(x/\eps+rk)}
\int_0^\infty ds e^{sQ}\int \frac{d\hat V(q)}{(2\pi)^d}
e^{is(k+p/2)\cdot q+i(x/\eps+rk)\cdot q}\\
&&\times\left[
\lambda(k+\frac p2-\frac q2)-\lambda(k+\frac p2+\frac q2)\right].
\end{eqnarray*}
This is further transformed to
\begin{eqnarray*}
&&\int dk e^{-ik\cdot y}\int\frac{d\hat V(p)}{(2\pi)^d}
e^{ip\cdot(x/\eps+rk)}
\int_0^\infty ds e^{sQ}\int \frac{d\hat V(q)}{(2\pi)^d}
e^{i(x/\eps+rk)\cdot q}
\int dy'\tilde\lambda(x,y')\\
&&\times\left[-e^{is(k-p/2)\cdot q}\left\{e^{iy'\cdot(k-p/2-q/2)}-
e^{iy'\cdot(k-p/2+q/2)}\right\}\right.\\
&&+\left.
e^{is(k+p/2)\cdot q}\left\{e^{iy'\cdot(k+p/2-q/2)}-
e^{iy'\cdot(k+p/2+q/2)}\right\}\right]\\
&&=\int \frac{d\hat V(p)}{(2\pi)^d}
e^{ip\cdot(x/\eps)}
\int_0^\infty ds e^{sQ}\int \frac{d\hat V(q)}{(2\pi)^d}
e^{ix/\eps\cdot q}
\tilde\lambda(x,y-rp-(r+s)q)\\
&&\times\left\{
-e^{-isp\cdot q/2-i(p+q)\cdot(y-rp-(r+s)q)/2}+
e^{-isp\cdot q/2-i(p-q)\cdot(y-rp-(r+s)q)/2}\right.\\
&&+\left.
e^{isp\cdot q/2+i(p-q)\cdot(y-rp-(r+s)q)/2}-
e^{isp\cdot q/2+i(p+q)\cdot(y-rp-(r+s)q)/2}\right\}.
\end{eqnarray*}
Therefore we obtain
\begin{eqnarray*}
&&|\tilde\lambda_2^\eps(x,y,\hat V)|\le
C\int_0^\infty dr e^{-\alpha r}\sup_{\hat V}\int \frac{|d\hat V(p)|}{(2\pi)^d}
\int_0^\infty ds e^{-\alpha s}\sup_{\hat V_1}
\int \frac{|d\hat V_1(q)|}{(2\pi)^d}\\
&&~~~~~~~~~~~~~~~~\times
|\tilde\lambda(x,y-rp-(r+s)q)|
+C\int_0^\infty dr e^{-\alpha r}|\widetilde{{\cal L}\lambda}(x,y)|,
\end{eqnarray*}
which is (\ref{eq:sublemma1}).

{\bf Proof of Lemma \ref{sublemma2}.}
The first inequality in (\ref{bdforlbar}) follows form the definition of
$\|\cdot\|_{\cal A}$, and the second is shown as follows. Let us define
\[
g(x,k)={\cal L}\lambda(x,k)=\int \frac{dp}{(2\pi)^d}
\hat {R}(\frac{k^2-p^2}{2},k-p)
\left[\lambda(x,p)-\lambda(x,k)\right].
\]
Taking the Fourier transform in $x$ and $k$ we obtain
\[
\hat g(\xi,y)=\int \frac{dxdkdpdy'}{(2\pi)^{2d}}e^{-i\xi\cdot x-ik\cdot y}
\hat { R}(\frac{k^2-p^2}{2},k-p)
\tilde\lambda(x,y')\left[e^{ip\cdot y'}-
e^{ik\cdot y'}\right].
\]
Integrating $x$ out we obtain
\begin{eqnarray*}
&&\hat g(\xi,y)=\int \frac{dkdpdy'}{(2\pi)^{2d}}e^{-ik\cdot y}
\hat { R}(\frac{k^2-p^2}{2},k-p)\hat\lambda(\xi,y')
\left[e^{ip\cdot y'}-
e^{ik\cdot y'}\right].
\end{eqnarray*}
We make a change of variables $k'=k-p$, $p'=(k+p)/2$ and drop the primes
to get
\begin{eqnarray*}
&&\hat g(\xi,y)=\int \frac{dkdpdy'}{(2\pi)^{2d}}
e^{-ip\cdot y-ik\cdot y/2+
ip\cdot y'}
\hat { R}(k\cdot p,k)\hat\lambda(\xi,y')\\
&&\times\left[e^{-ik\cdot y'/2}-
e^{ik\cdot y'2}\right]=
\int \frac{dkdpdy'ds}{(2\pi)^{2d}}
e^{-isk\cdot p-ip\cdot y-ik\cdot y/2+
ip\cdot y'}\\
&&\times
\tilde R(s,k)\hat\lambda(\xi,y)
\left[e^{-ik\cdot y'/2}-
e^{ik\cdot y'/2}\right].
\end{eqnarray*}
We may now integrate $p$ and $y'$ out to obtain
\begin{eqnarray*}
&&|\hat g(\xi,y)|\le 2\int \frac{dkds}{(2\pi)^d}|\hat R(s,k)|
|\hat\lambda(\xi,y+sk)|.
\end{eqnarray*}
Therefore we have
\[
\int d\xi dy|\hat g(\xi,y)|\le
2\|\tilde R(s,p)\|_{L_{p,s}^1}\int d\xi dy|\hat{\lambda}(\xi,y)|
\]
and thus (\ref{bdforlbar}) holds.

{\bf Proof of Lemma \ref{sublemma3}.} Clearly we have
\begin{equation}\label{lambda2linfty}
|I(x,y)|\le \frac{C}{\alpha^2}\|\tilde\lambda\|_{L_{x,y}^\infty}
\end{equation}
and thus it suffices to look at $|y|>2$.
We observe that
\begin{eqnarray*}
&&I_{r>S}(x,y)=\int_{S(y)}^\infty \!\!\! dr
e^{-\alpha r}\sup_{\hat V}\int \frac{|d\hat V(p)|}{(2\pi)^d}
\int_0^\infty \!\!\! ds
e^{-\alpha s}\sup_{\hat V_1}\int \frac{|d\hat V_1(q)|}{(2\pi)^d}
|\tilde\lambda(x,y-rp-(r+s)q)|\\
&&~~~~~~~~~~~~~~~~
\le \frac{C}{\alpha} e^{-\alpha S(y)}\|\tilde\lambda\|_{L_{x,y}^\infty}.
\end{eqnarray*}
Therefore we have
\[
\int dy\sup_{x}I_{r>S}(x,y)\le
{C}_{\alpha}\|\tilde\lambda\|_{L_{x,y}^\infty}.
\]
Now we look at $I_{r<S}$ and split it as:
\begin{eqnarray*}
&&I_{r<S}(x,y)=\int_{0}^{S(y)} \!\!\!dr
e^{-\alpha r}\sup_{\hat V}\int \frac{|d\hat V(p)|}{(2\pi)^d}
\int_0^\infty ds
e^{-\alpha s}\sup_{\hat V_1}\int \frac{|d\hat V_1(q)|}{(2\pi)^d}
|\tilde\lambda(x,y-rp-(r+s)q)|\\
&&~~~~~~~~~~~~~~~\le I_{r<S}^{s<S}(x,y)+
I_{r<S}^{s>S}(x,y).
\end{eqnarray*}
Observe that
\[
I_{r<S}^{s>S}(x,y)\le {C}_{\alpha} e^{-\alpha S(y)}
\|\tilde\lambda\|_{L_{x,y}^\infty}
\]
so that
\[
\int dy\sup_{x}I_{r<S}^{s>S}(x,y)\le {C}_{\alpha}
\|\tilde\lambda\|_{L_{x,y}^\infty}.
\]
It remains to bound $I_{r<S}^{s<S}$. Note that for $r,s\le
S(y)$, $|p|,|q|\le L$ and $\lambda$ in the Schwartz class we have
\begin{eqnarray*}
&&|\tilde\lambda(x,y-rp-(r+s)q)|\le
\sup_{|z-y|\le 2(r+s)L}|\tilde\lambda(x,z)|\le
\sup_{|z|\ge |y|/2}|\tilde\lambda(x,z)|
\le \frac{C_\lambda}{|y|^{5d}}.
\end{eqnarray*}
Therefore we have
\begin{eqnarray*}
&&\int_{|y|\ge 2} dy I_{r<S}^{s<S}(x,y)=\int_{|y|\ge 2} dy
 \int_{0}^{S(y)} dr
e^{-\alpha r}\sup_{\hat V}\int\frac{|d\hat V(p)|}{(2\pi)^d}
\int_0^{S(y)} ds
e^{-\alpha s}\sup_{\hat V_1}\int \frac{|d\hat V_1(q)|}{(2\pi)^d}\\
&&~~~~~~~~~~~~~~~~~~~~~~~~\times|\tilde\lambda(x,y-rp-(r+s)q)|
\le \int_{|y|\ge 2} dy\frac{C_\lambda}{|y|^{5d}}\le C_\lambda.
\end{eqnarray*}
This finishes the proof of Lemma \ref{sublemma3}.

\subsection{The tightness of the measures ${\cal P}_\eps$.}\label{sec:compact}

The process $W_\eps(t)$ generates a
probability measure $P_\eps$ on the space $C([0,T];X)$ with the space
$X$ defined as before $X=\left\{W\in{\cal S}':~\|W\|_{{\cal A}'}\le
C\right\}$. This family is tight.
\begin{lemma}\label{lemma-tight}
The family of measures $P_\eps$ is weakly compact.
\end{lemma}
{\bf Proof.}
We follow the corresponding proof of
Blankenship and Papanicolaou \cite{BP} for oscillatory ordinary
differential equations with random coefficients.
A theorem of Mitoma and Fouque \cite{Mitoma,Fouque}
implies that in order to verify tightness of the family $P_\eps$ it is
enough to check that for each $\lambda\in C^1([0,T],{\cal S}({\mathbb
R}^d\times{\mathbb R}^d))$ the family of measures ${\cal P}_\eps$ on
$C([0,T];{\mathbb R})$ generated by the random processes
$W_\lambda^\eps(t)=\langle W_\eps(t),\lambda\rangle$ is tight.
Tightness of ${\cal P}_\eps$ would follow from the following two conditions.
First, a Kolmogorov
moment condition \cite{billingsley1} in the form
\begin{equation}\label{eq:kolmo}
E^{P_\eps}\left\{\left|\langle
W,\lambda\rangle(t)-\langle W,\lambda\rangle(t_1)\right|^\gamma
\left|
\langle W,\lambda\rangle(t_1)-\langle W,\lambda\rangle(s)\right|^\gamma
\right\}\le
C_\lambda(t-s)^{1+\beta},~~~0\le s\le t\le T
\end{equation}
should hold with $\gamma>0$, $\beta>0$ and $C_\lambda$ independent of
$\eps$.
Second, we should have
\[
\lim_{R\to\infty}\limsup_{\eps\to 0}\hbox{Prob}^{{\cal P}_\eps}
\left\{\sup_{0\le t\le T}
\left|\langle W,\lambda\rangle(t)\right|>R\right\}=0.
\]
The second condition holds automatically in our case since
the process $W_\lambda^\eps(t)$ is uniformly
bounded for all $t>0$ and $\eps>0$.
In order to verify (\ref{eq:kolmo}),
note that we have
\[
\langle
W(t),\lambda\rangle=G_{\lambda_\eps}^\eps(t)-\sqrt{\eps}\langle
W,\lambda_1^\eps\rangle-\eps\langle W,\lambda_2^\eps\rangle+ \int_0^t
ds\langle W,\pdr{\lambda}{t}+k\cdot\nabla_x\lambda+{\cal
L}\lambda\rangle(s)+
\sqrt{\eps}\int_0^t ds\langle W,\zeta_\eps^\lambda\rangle(s).
\]
The uniform bound (\ref{zetabd}) on $\zeta_\eps^\lambda$ and the bounds
on $\|\lambda_{1,2}^\eps(t)\|_{\cal A}$ in Lemma
\ref{lemma1new} imply that it suffices to check (\ref{eq:kolmo}) for
\[
x_\eps(t)=G_{\lambda_\eps}^\eps(t)+
\int_0^t ds\langle W,\pdr{\lambda}{t}+k\cdot\nabla_x\lambda+{\cal
L}\lambda\rangle(s).
\]
We have
\begin{eqnarray*}
&&E\left\{\left.\left|x_\eps(t)-x_\eps(s)\right|^2\right|{\cal
F}_s\right\}\le 2E\left\{\left.\left|\int_s^t d\tau\langle
W,\pdr{\lambda}{t}+k\cdot\nabla_x\lambda+{\cal
L}\lambda\rangle(\tau)\right|^2\right|{\cal F}_s\right\}\\
&&+2E\left\{\left.
\left|G_{\lambda_\eps}^\eps(t)-G_{\lambda_\eps}^\eps(s)\right|^2
\right|{\cal F}_s\right\}\le
C(t-s)^2+2E\left\{\left.\langle G_{\lambda_\eps}^\eps\rangle(t)-
\langle G_{\lambda_\eps}^\eps\rangle(s)\right|{\cal F}_s\right\}.
\end{eqnarray*}
Here $\langle G_{\lambda_\eps}^\eps\rangle$ is the increasing process
associated with $G_{\lambda_\eps}^\eps$. We will now compute it
explicitly. First we obtain that
\[
\left.\frac d{dh}E_{W,\hat V,t}^{P_\eps}
\left\{\langle W,\lambda_\eps\rangle^2(t+h)\right\}\right|_{h=0}=
2\langle W,\lambda_\eps\rangle\langle
W,\pdr{\lambda}{t}+k\cdot\nabla_x\lambda_\eps+
\frac{1}{\sqrt{\eps}}{\cal K}[\hat V,x/\eps]\lambda_\eps\rangle+
\frac 1\eps Q\left[\langle W,\lambda_\eps\rangle^2\right]
\]
so that
\[
\langle W,\lambda_\eps\rangle^2(t)-\int_0^t\left(
2\langle W,\lambda_\eps\rangle(s)\langle
W,\pdr{\lambda}{t}+k\cdot\nabla_x\lambda_\eps+
\frac{1}{\sqrt{\eps}}{\cal K}[\hat V,x/\eps]\lambda_\eps\rangle(s)+
\frac 1\eps Q\left[\langle W,\lambda_\eps\rangle^2\right](s)\right)ds
\]
is a martingale.
Therefore we have
\begin{eqnarray*}
&&\langle G_{\lambda_\eps}^\eps(t)\rangle=
\int_0^tds\left[\frac 1\eps Q[\langle W,\lambda_\eps\rangle^2]-\frac 2\eps
\langle W,\lambda_\eps\rangle\langle W,Q\lambda_\eps\rangle\right](s)\\
&&~~~~~~~~~~~
=\int_0^t ds \left(Q\left[\langle W,\lambda_1^\eps\rangle^2\right]-
\langle W,\lambda_1^\eps\rangle\langle W,Q\lambda_1^\eps\rangle(s)\right)+
\sqrt{\eps }\int_0^tds H_\eps(s)
\end{eqnarray*}
with
\begin{eqnarray*}
&&H_\eps=2\sqrt{\eps}\left(Q[\langle W,\lambda_1^\eps\rangle
\langle W,\lambda_2^\eps\rangle]-\langle W,\lambda_1^\eps\rangle
\langle W,Q\lambda_2^\eps\rangle-\langle W,\lambda_2^\eps\rangle
\langle W,Q\lambda_1^\eps\rangle\right)\\
&&~~~~~~~~~~~
+\eps\left(Q[\langle W,\lambda_2^\eps\rangle^2]
-2\langle W,\lambda_2^\eps\rangle\langle W,Q\lambda_2^\eps\rangle\right).
\end{eqnarray*}
Lemma \ref{lemma1new} and the boundedness of $Q$ on $L^\infty({\cal V})$
imply that $|H_\eps(s)|\le C$ for all $V\in{\cal V}$.
This yields
\[
E\left\{\left.\langle G_{\lambda_\eps}^\eps\rangle(t)-\langle
G_{\lambda_\eps}^\eps\rangle(s)\right|{\cal F}_s\right\} \le C(t-s)
\]
and hence
\begin{eqnarray*}
E\left\{\left.\left|x_\eps(t)-x_\eps(s)\right|^2\right|{\cal
F}_s\right\}\le C(t-s).
\end{eqnarray*}
In order to obtain (\ref{eq:kolmo}) we note that
\begin{eqnarray*}
&&E^{P_\eps}\left\{ \left|x_\eps(t)-x_\eps(t_1)\right|^\gamma
\left|x_\eps(t_1)-x_\eps(s)\right|^\gamma\right\}\\ &&= E^{P_\eps}
\left\{E^{P_\eps}\left\{\left.\left|x_\eps(t)-x_\eps(t_1)\right|^\gamma
\right|{\cal F}_{t_1}
\right\}\left|x_\eps(t_1)-x_\eps(s)\right|^\gamma\right\}\\ &&\le
E^{P_\eps}
\left\{\left[E^{P_\eps}\left\{
\left.\left|x_\eps(t)-x_\eps(t_1)\right|^2\right|{\cal
F}_{t_1}
\right\}\right]^{\gamma/2}\left|x_\eps(t_1)-x_\eps(s)\right|^\gamma\right\}\\
&&\le
C(t-t_1)^{\gamma/2}E^{P_\eps}\left\{
\left|x_\eps(t_1)-x_\eps(s)\right|^\gamma\right\}\le
C(t-t_1)^{\gamma/2}E^{P_\eps}\left\{E^{P_\eps}\left\{
\left|x_\eps(t_1)-x_\eps(s)\right|^\gamma|{\cal F}_s\right\}\right\}\\
&&\le C(t-t_1)^{\gamma/2}E^{P_\eps}\left\{
\left.\left[E^{P_\eps}\left\{
\left|x_\eps(t_1)-x_\eps(s)\right|^2\right|{\cal
F}_{s}\right\}\right]^{\gamma/2}\right\}
\le C(t-t_1)^{\gamma/2}(t_1-s)^{\gamma/2}\\
&&\le C(t-s)^\gamma.
\end{eqnarray*}
Choosing now $\gamma>1$ we get (\ref{eq:kolmo}) which finishes the
proof of Lemma \ref{lemma-tight}. This also finishes the proof of
Theorem \ref{thm1}.

\section{Conclusions}

We have presented a proof of the transport limit (\ref{rte}) for the
Schr\"odinger equation with a time-dependent random potential. Our
proof is relatively simple and does not involve infinite Neumann
expansions because it relies on the Markovian property of the
potential, which allows us to construct approximate martingales and to
show weak compactness of the family of probability measures $P_\eps$
generated by the dynamics (\ref{eq:wigeqnew}) of the Wigner transform
on $C([0,T];{\cal A}')$. However, we only show convergence for the
average Wigner distribution, which is the the first moment of
$P_\eps$.  We do not have a rigorous convergence result for the higher
moments of the Wigner distribution and thus are not able to fully
characterize the set of accumulation points of the family $P_\eps$,
although we believe that the limit measure $P$ is unique, based on the
formal analysis in
\cite{PRS} of a similar problem in the white noise limit.

\begin{appendix}

\section{The formal perturbation expansion}\label{sec:append}

We present the formal derivation of the transport equation for the
limit Wigner distribution $W(t,x,k)$ similar to the one in
\cite{RPK-WM} for a time-independent potential.  Recall that the Cauchy problem
for the Wigner distribution is
\begin{eqnarray}\label{eq:wigeqnew-app}
&&\pdr{W_\eps}{t}+k\cdot\nabla_xW_\eps=\frac 1{i\sqrt{\eps}}
\int \frac{d\tilde V(t/\eps,p)}{(2\pi)^d} e^{ip\cdot x/\eps}
\left[W_\eps(t,x,k-\frac{p}{2})-W_\eps(t,x,k+\frac{p}{2})\right]\\
&&W_\eps(0,x,k)=W_\eps^0(x,k).\nonumber
\end{eqnarray}
Here $W_\eps^0$ is the Wigner distribution of the family
$\phi_\eps^0(x)$, the initial data for (\ref{eq:schroed}).  
We will construct a formal perturbation expansion for $W_\eps$ and
derive the transport equation for the average $\overline
W(t,x,k)$. 
We seek an expansion of $W_\eps$ with multiple scales 
\begin{equation}\label{eq:wexp}
W_\eps(t,x,k)=W^{(0)}(t,x,k)+\sqrt{\eps}W^{(1)}(t,\frac t\eps,x,\frac
x\eps,k)+{\eps}W^{(2)}(t,\frac t\eps,x,\frac x\eps,k)+\dots
\end{equation}
and introduce the fast time and spatial variables $\tau=t/\eps$,
$z=x/\eps$. We assume that the leading term $W^{(0)}$ is deterministic
and independent of the fast scale variables. We insert (\ref{eq:wexp})
into (\ref{eq:wigeqnew-app}) and obtain at the order $O(1/\sqrt{\eps})$:
\begin{eqnarray*}
\partial_\tau W^{(1)} + k\cdot\nabla_z W^{(1)}
={\cal K}[\hat V,z]W^{(0)}.
\end{eqnarray*}
Then $W^{(1)}$ has the form
\begin{equation}\label{eq:w1}
W^{(1)}(t,\tau,x,z,k)=\frac 1i\int \frac{dp d\omega
e^{ip\cdot z+i\omega s}}{(2\pi)^{d+1}} \frac{\hat
V(\omega,p)}{i\omega+ip\cdot k+\delta}
\left[W^{(0)}(x,k-\frac{p}{2})- W^{(0)}(x,k+\frac{p}{2})\right].
\end{equation}
Here $\delta\ll 1$ is a regularization parameter that we will send to
zero at the end of the calculation and $\hat V$ denotes the Fourier
transform in time:
\[
\hat V(\omega,p)=\int_{\mathbb R} ds e^{-i\omega s}\tilde V(s,p).
\]
The term of order $O(1)$ in
(\ref{eq:wigeqnew}) gives
\[
\pdr{W^{(0)}}{t}+\pdr{W^{(2)}}{\tau} + k\cdot\nabla_x
W^{(0)}+k\cdot\nabla_z W^{(2)}
=\frac 1i\int \frac{dp\tilde V(\tau,p)e^{ip\cdot z}}{(2\pi)^d}
\left[W^{(0)}(k-\frac p2)-W^{(0)}(k+\frac p2)\right].
\]
We average the above equation assuming formally that
$E\left\{\pdr{W^{(2)}}{\tau}+k\cdot\nabla_z W^{(2)}
\right\}=0$. This gives an equation for the leading order
term $W^{(0)}$:
\begin{eqnarray}\label{eq:w0}
\pdr{W^{(0)}}{t}+ k\cdot\nabla_x W^{(0)}=E\left\{\frac 1i
\int \frac{dp\tilde V(\tau,p)e^{ip\cdot z}}{(2\pi)^d}
\left[W^{(1)}(x,z,k-\frac p2)-W^{(1)}(x,z,k+\frac p2)\right]\right\}.
\end{eqnarray}
The average on the right side of (\ref{eq:w0}) may be computed
explicitly using (\ref{eq:w1}) and spatial homogeneity
(\ref{powerspectrum}):
\begin{eqnarray*}
&&E\left\{\frac 1i
\int \frac{dp\tilde V(\tau,p)e^{ip\cdot z}}{(2\pi)^d}
\left[W^{(1)}(x,k-\frac p2)-W^{(1)}(x,z,k+\frac p2)\right]\right\}\\
&&\to\int\frac{dp}{(2\pi)^d}\hat R(\frac{p^2-k^2}{2},p-k)\left[W^{(0)}(p)-
W^{(0)}(k)\right]
\end{eqnarray*}
as $\delta\to 0$. Here $\hat R$ is the Fourier transform of $\tilde R$
in time:
\[
\hat R(\omega,p)=\int_{\mathbb R}ds e^{-i\omega s}\tilde R(s,p).
\]
Therefore we obtain the transport equation for the leading order term
$W^{(0)}$:
\begin{equation}\label{eq:treqformal-app}
\pdr{W^{(0)}}{t}+ k\cdot\nabla_x W^{(0)}=
\int\frac{dp}{(2\pi)^d}\hat R(\frac{p^2-k^2}{2},p-k)\left[W^{(0)}(p)-
W^{(0)}(k)\right].
\end{equation}

The formal asymptotic expansion (\ref{eq:wexp}) may not be justified
but the final equation (\ref{eq:treqformal}) for the expectation of
the limit Wigner distribution $E\left\{W(t,x,k)\right\}$ is
correct. Moreover, the test functions that we used in our proof of
Theorem \ref{thm1} are based on the formal expressions for $W^{(1)}$
and $W^{(2)}$. The role of the regularization parameter $\delta$ is
played by the spectral gap of the generator $Q$ because of the bound
(\ref{eq:expdecaynew}).

\end{appendix}

\end{document}